\documentclass[pra,showpacs,superscriptaddress,groupedaddress]{revtex4}
\usepackage{ae}
\usepackage[T1]{fontenc}
\usepackage[ansinew]{inputenc}
\usepackage{amsmath}
\usepackage{amssymb}
\usepackage[caption=false]{subfig}
\usepackage{multirow}
\usepackage{array}
\usepackage[]{graphicx}
\usepackage{wrapfig}
\usepackage{apalike}
\usepackage{bbold}

\usepackage{color}
\usepackage[colorlinks]{hyperref}
\usepackage{lscape}
\begin{document}
\title{Maximal qubit violation of n-local inequalities in quantum network}
\author{Amit Kundu}
\email{amit8967@gmail.com}
\affiliation{Department of Applied Mathematics, University of Calcutta, 92 A.P.C Road, Kolkata- 700009, India}
\author{Mostak Kamal Molla}
\email{mostakkamal@gmail.com}
\affiliation{Department of Applied Mathematics, University of Calcutta, 92 A.P.C Road, Kolkata- 700009, India}
\author{Indrani Chattopadhyay}
\email{icappmath@caluniv.ac.in}
\affiliation{Department of Applied Mathematics, University of Calcutta, 92 A.P.C Road, Kolkata- 700009, India}
\author{Debasis Sarkar}
\email{dsarkar1x@gmail.com, dsappmath@caluniv.ac.in}
\affiliation{Department of Applied Mathematics, University of Calcutta, 92 A.P.C Road, Kolkata- 700009, India}
\begin{abstract}
Source independent quantum networks are considered as a natural generalization to the Bell scenario where we investigate the nonlocal properties of quantum states distributed and measured in a network. Considering the simplest network of entanglement swapping, recently Gisin et. al. and Andreoli et. al. independently provided a systematic characterization of the set of quantum states leading to violation of the so-called 'bilocality' inequality. In this work, we consider the complexities in the quantum networks with an arbitrary number of parties distributed in chain-shaped and star-shaped networks. We derive the maximal violation of the 'n-local' inequality that can be achieved by arbitrary two-qubit states for such chain and star-shaped networks. This would further provide us deeper understanding of quantum correlations in complex structures.
\end{abstract}
\maketitle
\section{Introduction}
Since its foundation in the early decades of the last century, quantum theory has been elevated to the status of the most precisely tested and successful theory in the history of science. Yet, many of its consequences have puzzled most of the scientists. Many counter-intuitive features of quantum theory like quantum entanglement\cite{2}, plays a central role in the foundation of the theory as well as in the development of the quantum information and computation theory\cite{18}. Another important feature is non-locality, shown initially through Bell's theorem that expresses the nature of quantum correlations exist between distant particles/parties of a composite quantum system, thus precluding its explanation by any local hidden variable(LHV) model.\\

Not surprisingly many generalizations of Bell's theorem\cite{3} have been pursued over the years. Immediate natural generalizations of this simple scenario include more measurements per party\cite{5}, sequential measurements\cite{6}, several measurement outcomes\cite{7}, increasing number of parties\cite{9, 10} and stronger notions of quantum nonlocality\cite{11, 12}. A common feature of all those generalizations is the correlations between the parties assumed to be mediated by the same source of states. However, in a quantum network scenario, the correlations between the distance nodes are not restricted by a single source, rather it is assumed that there are many independent sources distribute entanglement in a non-trivial way across the whole network and thus generates a strong correlation among several nodes. The simplest network scenario is provided by entanglement swapping\cite{13}, where two distant parties,say, Alice and Charlie, share entangled states with a central node Bob. Upon measuring in an entangled basis and conditioning on his outcomes, Bob can generate entanglement and non-local correlations among the two other distant parties even though they had no direct interactions prior to the measurement. To compare classical and quantum correlations in this scenario, it is natural to consider classical models consisting of two independent hidden variables, and we come to a notion of bilocality assumption\cite{16,17}. The bilocality scenario and generalizations to networks with an increasing number($n$) of independent sources of states, called $n-local$ scenario\cite{14,15,16,17,18}, allows us the emergence of a new kind of non-local correlations.\\

The main goal of this work is to provide the maximal violation of the $n-local$ inequalities for the $n-local$ scenario using arbitrary qubit states shared between two parties from different independent sources like the bilocal inequalities given by Andreoli et. al. and Gisin et. al. independently\cite{1, 14}. Here, we consider a chain-shaped network and a star-shaped network\cite{4, 15}.

\section{Characterizing Bilocal Correlation}
\begin{figure}[h]
\centering
\includegraphics[width=15cm, height=6cm]{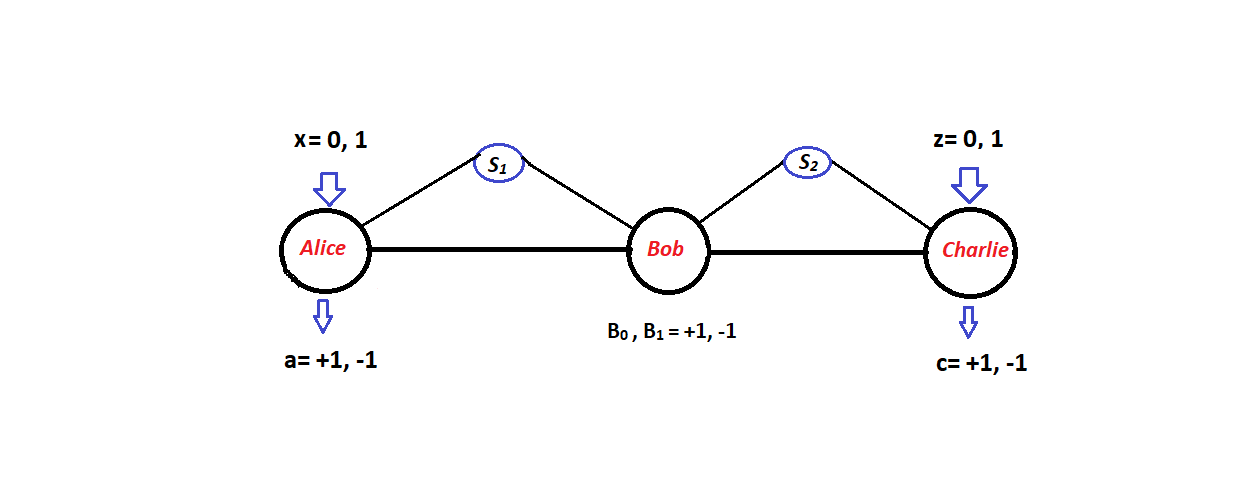}
\caption{Bilocal Network scenario}
\label{fig:bilocal1}
\end{figure}

In this section, we will briefly describe the basic structure of the bilocal network (see Figure \ref{fig:bilocal1}). Here, Alice and Bob are sharing a quantum state produced by the source $S_{1}$,  and Bob and Charlie are sharing a quantum state produced by another source  $S_{2}$.  One can generalize this notion to networks with an increasing number($n$) of independent sources of states, the so-called $n-local$ scenario, allow one to study a new kind of correlations.
For instances, correlations that appear classical, according to the usual local hidden variable (LHV) models can also display non-classicality if the independence of the sources is taken into account. More specifically, a bilocal correlation consists of three distant observers Alice, Bob and Charlie correlated by two independent sources of states. In the quantum case, Bob shares two pairs of entangled particles, one with Alice and another with Charlie. Now Alice, Bob and Charlie perform measurements $X$, $Y$, and $Z$ with outcomes respectively $a$, $b$, and $c$. The special feature of this scenario is that Bob is carrying two different particles produced by two independent sources with different hidden variables, say, ($\lambda_{1}$, $\lambda_{2}$). Any probability distribution compatible with the bilocality assumption, i.e., independence of the sources can be decomposed as,
\begin{equation}
p(a,b,c|x,y,z) = \int d\lambda_{1} d\lambda_{2}p(\lambda_{1})p(\lambda_{2})p(a|x, \lambda_{1})p(b|y,\lambda_{1}, \lambda_{2})p(c|z, \lambda_{2})
\end{equation}

with binary inputs of Alice and Charlie $x=0, 1$ and $z= 0, 1$ and with outputs $\pm 1$ in each case. The middle party, Bob always performs the same measurement with input $y$, with four possible outcomes, e.g., the BSM(Bell State Measurement), denoting Bob's outcome by two bits $B_{0}$ = $\pm 1$ and $B_{1} = \pm 1$. It follows that any bilocal hidden variable model described by the previous equation must full-fill the following bilocal inequality,
\begin{equation}
S_{biloc} = \sqrt{|I|} + \sqrt{|J|} \leq 2
\end{equation}
where 
$$I = \langle(A_{0} + A_{1})B_{0}(C_{0} + C_{1})\rangle$$
$$J = \langle(A_{0} - A_{1})B_{1}(C_{0} - C_{1})\rangle$$
and $\langle |\rangle$ denotes the expectation value of multiple experimental runs.

\section{Motivation}
\begin{figure}[h]
\centering
\includegraphics[width=15cm, height=6cm]{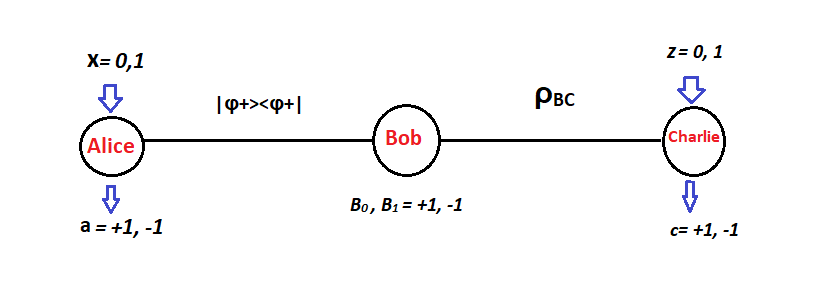}
\caption{Alice and Bob are sharing $|\phi^+\rangle\langle\phi^+|$ and a generic two-qubit state is shared between Bob and Charlie.}
\label{fig:bilocal2}
\end{figure}

From the previous section, we learn that any correlation distributed between Alice, Bob and Charlie will be said non- bilocal if it violates the inequality (2). In\cite{1}, Gisin et. al. showed that all the pure entangled states violate this bilocal inequality. They also showed that for mixed states distributed over the three parties from two different sources there is a relation between bilocal inequality violation and CHSH inequality violation for each independent source. They proved the following relation,
\begin{equation}
S_{biloc}^{max} \leq \sqrt{S_{AB(CHSH)}^{max}S_{BC(CHSH)}^{max}}
\end{equation}
So, the violation of bilocal inequality implies that either $\rho_{AB}$, or $\rho_{BC}$, or both must violate CHSH Inequality\cite{3}, where $\rho_{AB}$ and $\rho_{BC}$ be two mixed states shared between Alice, Bob and Bob, Charlie respectively with the generic form,
$$\rho_{AB} = \frac{1}{4}(\mathbb{1}+\vec{m_{A}}.\vec{\sigma}\otimes\mathbb{1}+\mathbb{1}\otimes\vec{\sigma}.\vec{m_{B}}+\sum_{mn}t_{mn}\sigma_{m}\otimes\sigma_{n})$$
and 
$$\rho_{BC} = \frac{1}{4}(\mathbb{1}+\vec{m_{B}}.\vec{\sigma}\otimes\mathbb{1}+\mathbb{1}\otimes\vec{\sigma}.\vec{m_{C}}+\sum_{mn}t_{mn}\sigma_{m}\otimes\sigma_{n})$$
Here, the vector $\vec{m_{A}}$($\vec{m_{B}}$) represents the Bloch vector of Alice's (Bob's) reduced state, while $t_{mn}$ (with $m, n \in {x, y, z}$) is the terms in the correlation matrix, and similarly for $\rho_{BC}$.
Maximizing $S_{biloc}$, they proved that the minimum criteria for bilocal violation are either $\rho_{AB}$ or $\rho_{BC}$ be non-local.
In support of this statement, F. Andreoli et. al., took the state $|\phi^+\rangle\langle\phi^+|, ~~|\phi^+\rangle = \frac{|00 \rangle + |11\rangle }{\sqrt{2}} $(one of the  four Bell states) between Alice and Bob and a generic two qubit state between Bob and Charlie and showed that the generic state while satisfying CHSH inequality, $|\phi^+\rangle\langle\phi^+| \otimes \rho_{BC}$, violates bilocal inequality(Figure {\ref{fig:bilocal2}}).
Observing these results, one should ask what would be the corresponding results if one chooses more complicated quantum network scenario with such generic states. We have tried to answer this question in two specific complicated network scenarios.
\section{n- Local scenario}
\begin{figure}[h]
\centering
\includegraphics[width=16cm, height=5cm]{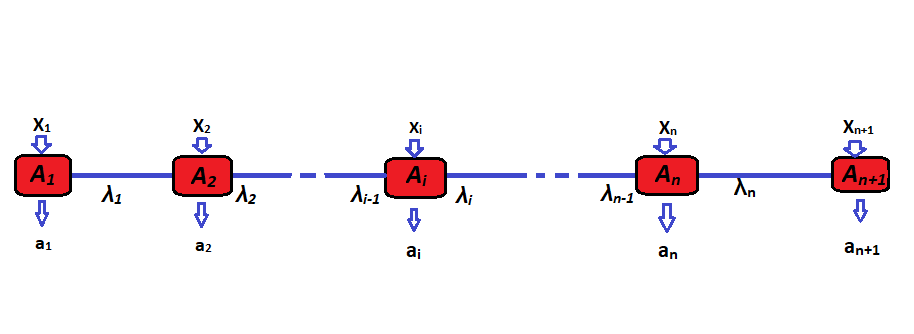}
\caption{n-local chain shaped Network scenario}
\label{fig:bilocal3}
\end{figure}

We first consider the $n-local$ chain-shaped scenario, depicted in figure {\ref{fig:bilocal3}}. There are $n$ sources $S_{i}$ ($i$=1,....,$n$) and $n+1$ parties $A_{i}$ ($i$=1,.....,$n+1$) arranged in a linear chain such that any two neighbouring parties share a common source. For each $i (i=1,...,n+1)$, party $A_{i}$ can perform two dichotomic measurements $x_{i} = x_{i}^k(k=1,2)$ where $x_{i}^k \in \{ 0,1 \}$ on the respective subsystems they have received and obtain outcomes $a_{i}$ = $a_{i}^j($j = 1,2$)$ where $a_{i}^j\in \{ 0,1 \}$. Except the extreme two parties $A_{1}$ and $A_{n+1}$, each of the remaining $n-1$ parties who receive two subsystems, the measurement will be joint measurement operating on both the subsystems simultaneously\cite{15}. In this scenario, Bell's locality assumption takes the form:
\begin{equation}
P(a_{1},...,a_{n+1|x_{1},...,x_{n+1}}) = \int d\lambda \rho(\lambda)\prod_{i=1}^{n+1}P(a_{i}|x_{i},\lambda)
\end{equation}
Here $\lambda$ is the joint hidden state. For each party $A_{i}$, the input $x_{i}$ and the corresponding output $a_{i}$ is determined by the local probability distributions $P(a_{i}|x_{i}, \lambda)$. The hidden state follows the distribution $\rho(\lambda)$, satisfying the normalization condition $\int\rho(\lambda)d\lambda = 1$, where we assume that the measurement choices of each party are independent of $\lambda$.
Now, suppose each of the sources $S_{i}$ be characterized by the hidden state $\lambda_{i}$, Moreover, we assume that for each $i \in \{ 2,...,n \}$, the output of the party $A_{i}$ depend on the hidden states $\lambda_{i-1}, \lambda_{i}$ and its input $x_{i}$, whereas for $A_{1}$ and $A_{n+1}$ the outputs depend on  their corresponding inputs and $\lambda_{1}$ and $\lambda_{n}$, respectively. So, we can write,

$$P(a_{1},   ,a_{n+1|x_{1},...,x_{n+1}}) = \int d\lambda_{1}...\int d\lambda_{n}\rho(\lambda_{1},...,\lambda_{n})P(a_{1}|x_{1})$$ $$
\times\prod_{i=2}^n P(a_{i}|x_{i},\lambda_{i-1}\lambda_{i})
P(a_{n}|x_{n+1},\lambda_{n}).$$
Without any further assumption this equation can jump to usual Bell local equation, when all the different hidden states will be the same. So, for $n-local$ assumption, the distribution of the hidden states with independent sources, we can factorize the probability distribution as;
$$\rho(\lambda_{1},...,\lambda_{n}) = \prod_{i}\rho_{i}(\lambda_{i})$$.
The independence of the sources and probability distributions as factorized above, together define the assumption of n-locality. To study whether a given correlation is n-local or not there is a non-linear inequality exactly same like bilocal inequality which is referred to as $n-local$ inequalities. Here, we only consider the case, where the two extreme parties have binary inputs and outputs and with one input and four outputs for the remaining $n-1$ parties (i.e., $P^{14}$ scenario). Another picture is each party has two inputs and two outputs($P^{22} Scenario$). We observe that for n=2, the scenario exactly like bilocal\cite{17} for which Gisin et. al.\cite{1} proved the inequality in equation (3). Here, we try to extend the result for this $n-local$ network with $P^{14}$ scenario.

To check the $n-local$ correlation we have the following inequality;
\begin{equation}
\sqrt{|I|}+\sqrt{|J|} \leq 2
\end{equation} where,
\begin{equation}
I = \sum_{x_{1},x_{n+1} = 0,1}\langle A_{1,x_{1}}A_{2}^0...A_{n}^0A_{n+1,x_{n+1}}\rangle
\end{equation}
\begin{equation}
J = \sum_{x_{1},x_{n+1} = 0,1}(-1)^{x_{1}+x_{n+1}}\langle A_{1,x_{1}}A_{2}^1...A_{n}^1A_{n+1,x_{n+1}}\rangle
\end{equation}
and $A^0$, $A^1$ be the joint measurements of the middle parties with outputs $A^0 = \pm 1$ and $A^1 = \pm 1$.
From the above inequality, one would be able to observe that if a correlation violates the inequality then the correlation is non- n-local and also it can be seen \cite{15} that for quantum correlations the inequality is violated. So, for this $n-local$ scenario we find the minimum number of sources which is CHSH non-local and violate the n-local inequality. And later we extend this result to star-shaped $n-local$ scenario also, where there is a central node connected with $n$ number of parties with $n$ independent sources.

\section{Violation criteria for linear chain network}
In \cite{1}, for bilocal network, sharing two generic mixed states $\rho_{AB}$ and $\rho_{BC}$ in Sec:III, provided by two independent sources, the maximum value of the left-hand side of the bilocality inequality (5) is given by, 
\begin{equation}
S_{biloc}^{max} = 2\sqrt{\sqrt{\Lambda_{1}^A\gamma_{1}^C} + \sqrt{\Lambda_{2}^A\gamma_{2}^C}}
\end{equation} 
where, $\Lambda_{1} \geq \Lambda_{2} \geq \Lambda_{3} \geq 0$ are the three eigenvalues of the matrix $R^{AB}=t^{AB\dag}t^{BC}$ formed by the correlation matrix $t^{AB} = U^{AB}R^{AB}$, denoting $U^{AB}$ is a unitary matrix. Similarly, $\gamma_{1} \geq \gamma_{2} \geq \gamma_{3} \geq 0$ for $R^{BC}$.

Now, maximal CHSH value for $\rho_{AB}$ is $S_{AB}^{max} = 2\sqrt{\Lambda_{1}+\Lambda_{2}}$ and similarly for $\rho_{BC}$. So, we have, $S_{BC}^{max} = 2\sqrt{\gamma_{1}+\gamma_{2}}$. It directly follows $S_{biloc}^{max}\leq \sqrt{S_{AB}^{max}S_{BC}^{max}}$, which proves that the violation of the bilocality inequality implies that either $\rho_{AB}$ or $\rho_{BC}$ or both must violate CHSH inequality. 
In[5], they showed, if Alice and Bob share a maximally entangled state(Bell State) while Bob and Charlie share a generic quantum state then bilocal inequality violation can be easily obtained without CHSH violation of $\rho_{BC}$. $$S_{biloc}^{max}(|\phi^+\rangle\langle\phi^+| \otimes \rho_{BC}) = \sqrt{\sqrt{\gamma_{1}} + \sqrt{\gamma_{2}}} \geq \sqrt{\gamma_{1} + \gamma_{2}} = S_{BC}^{max(CHSH)}$$

\begin{figure}[h]
\centering
\includegraphics[width=14cm, height=8cm]{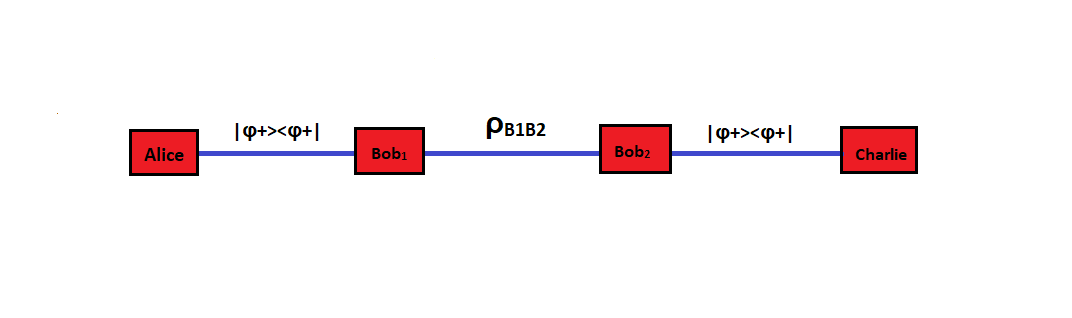}
\caption{3-local chain shaped Network scenario}
\label{fig:bilocal4}
\end{figure}

Considering a little bit more complex network than above, we introduce first one more party (instead of three, now four), called Alice, Bob1, Bob2, Charlie as in the figure {\ref{fig:bilocal4}}. Here, the number of independent sources are three and we can call this situation as 3-local (trilocal) scenario, a special case of n-local scenario. The middle two parties, i.e., Bob1 and Bob2 again will do Bell State measurements with one input and four outputs denoted by $B_{1}^0 = \pm1, B_{1}^1 = \pm1, B_{2}^0 = \pm1, B_{2}^1 = \pm1$. The inequality for this 3-local scenario is,
$$\sqrt{|I|}+\sqrt{|J|} \leq 1$$ where,
 $$I = \frac{1}{4}\langle(A_{0}+A_{1})B_{1}^0B_{2}^0(C_{0}+C_{1})$$
and 
 $$J = \frac{1}{4}\langle(A_{0}-A_{1})B_{1}^1B_{2}^1(C_{0}-C_{1})$$

Now, we take $\rho_{AB_{1}} = |\phi^+\rangle\langle\phi^+|$, $\rho_{B_{2}C} = |\phi^+\rangle\langle\phi^+|$ and $\rho_{B_{1}B_{2}}$ as a generic quantum state as in figure {\ref{fig:bilocal4}}. If we choose measurement choices of Alice and Charlie be any Bloch vector $(sin\alpha, 0, cos\alpha)$, $(sin\alpha^\prime, 0, cos\alpha^\prime)$ and $(sin\theta, 0, cos\theta)$, $(sin\theta^\prime, 0, cos\theta^\prime)$ and $B_{1}^0 = B_{2}^0 = \sigma_{z}\otimes \sigma_{z}$ and $B_{1}^1 = B_{2}^1 = \sigma_{x}\otimes \sigma_{x}$, we can determine the bilocal value for $\rho_{AB_{1}}\otimes\rho_{B_{1}B_{2}}\otimes\rho_{B_{2}C}$ with this measurements settings. After maximizing the bilocal value we get $S_{3-local}^{max} = \sqrt{\sqrt{\Lambda_{1}}+\sqrt{\Lambda_{2}}}$, where $\Lambda_{1}$ and $\Lambda_{2}$ are the two maximum eigenvalues of the matrix $R^{B_{1}B_{2}} = (t^{B_{1}B_{2}} )^\dag t^{B1B2}$. So, it can be easily seen that to violate 3-local inequality $\rho_{B_{1}B_{2}}$ need not be non-local.

Similarly if we take $\rho_{AB_{1}} = |\phi^+\rangle\langle\phi^+|$ with $\rho_{B_{1}B_{2}}$ and $\rho_{B_{2}C}$ be generic states, we get the bilocal value, $\sqrt{\sqrt{\Lambda_{1}\gamma_{1}}+\sqrt{\Lambda_{2}\gamma_{2}}}$, where $\Lambda_{1}$ and $\Lambda_{2}$ are two maximum eigenvalues of the matrix $R^{B_{1}B_{2}} =( t^{B_{1}B_{2} })^\dag t^{B_{1}B_{2}}$ and similarly  $\gamma_{1}$ and $\gamma_{2}$ for $R^{B_{2}C}$ .

Now, we take all the three states in generic from and calculate the bilocal value for arbitrary two measurement choices of Alice and Charlie. The quantity $I$ can be expressed as follows:
\begin{equation}
I = \frac{1}{4}Tr[(a+a^\prime).\vec{\sigma} \otimes \sigma_{z}\otimes \sigma_{z} \otimes \sigma_{z}\otimes \sigma_{z} \otimes (c+c^\prime). \vec{\sigma} \rho_{AB_{1}}\otimes\rho_{B_{1}B_{2}}\otimes\rho_{B_{2}C}]
\end{equation}
$$=\frac{1}{4}Tr[(a+a^\prime).\vec{\sigma} \otimes \sigma_{z}\rho_{AB_{1}}]Tr[\sigma_{z} \otimes \sigma_{z}\rho_{B_{1}B_{2}}]Tr[\sigma_{z} \otimes (c+c^\prime). \vec{\sigma}\rho_{B_{2}C}] $$
$$= \frac{1}{4}\sum_{i = x,y,z}(a_{i}+a_{i}^\prime)t_{iz}^{AB_{1}}.t_{zz}^{B_{1}B_{2}}.\sum_{j=x,y,z}t_{zj}^{B_{2}C}(c_{j}+c_{j}^\prime) $$
And similarly,
\begin{equation}
J = \frac{1}{4}Tr[(a-a^\prime).\vec{\sigma} \otimes \sigma_{x}\otimes \sigma_{x} \otimes \sigma_{x}\otimes \sigma_{x} \otimes (c-c^\prime). \vec{\sigma} \rho_{AB_{1}}\otimes\rho_{B_{1}B_{2}}\otimes\rho_{B_{2}C}]
\end{equation}
$$=\frac{1}{4}Tr[(a-a^\prime).\vec{\sigma} \otimes \sigma_{x}\rho_{AB_{1}}]Tr[\sigma_{x} \otimes \sigma_{x}\rho_{B_{1}B_{2}}]Tr[\sigma_{x} \otimes (c-c^\prime). \vec{\sigma}\rho_{B_{2}C}] $$
$$= \frac{1}{4}\sum_{i = x,y,z}(a_{i}-a_{i}^\prime)t_{ix}^{AB_{1}}.t_{xx}^{B_{1}B_{2}}.\sum_{j=x,y,z}t_{xj}^{B_{2}C}(c_{j}-c_{j}^\prime) $$

As said earlier, the correlation matrix can be written in the polar decomposition form \cite{1}. Here, we can denote the eigenvalues of the matrix $R^{B_{1}B_{2}}$ as, $\beta_{1}\geq\beta_{2}\geq\beta_{3}$.
Next, to prove our aim, we maximize $S_{3-local}$ with respect to the Bloch vectors $\vec{a}$, $\vec{a^\prime}$, $\vec{c}$ and $\vec{c^\prime}$. Considering the form of the vector $\vec{a}=(sin\alpha, 0, cos\alpha)$, $\vec{a^\prime}=(sin\alpha^\prime, 0, cos\alpha^\prime)$ and $\vec{c}=(sin\theta, 0, cos\theta)$, $\vec{c^\prime}=(sin\theta^\prime, 0, cos\theta^\prime),$ we can easily find the maximum value by imposing the conditions $\partial_{\alpha}S = 0$, $\partial_{\alpha^\prime}S = 0$, $\partial_{\theta}S = 0$ and $\partial_{\theta^\prime}S = 0$
and obtain [1],
\begin{equation}
S_{3-local}^{max} = \sqrt{\sqrt{\Lambda_{1}\gamma_{1}\beta_{1}}+\sqrt{\Lambda_{2}\gamma_{2}\beta_{2}}}
\end{equation}
Consequently, the states $\rho_{AB_{1}}$, $\rho_{B_{1}B_{2}}$ and $\rho_{B_{2}C}$ can violate the 3-local inequality if and only if $\sqrt{\Lambda_{1}\gamma_{1}\beta_{1}}+\sqrt{\Lambda_{2}\gamma_{2}\beta_{2}} > 1$.

For this inequality a criterion analogous to the Horodecki criteria\cite{2} could be found easily by applying Cauchy- Schwarz inequality:
\begin{equation}
S_{3-local}^2 < S_{AB_{1}}^{max(CHSH)}S_{B_{1}B_{2}}^{max(CHSH)}S_{B_{2}C}^{max(CHSH)}
\end{equation}

We can easily extend this result for n-local scenario. If we take $\rho_{AB_{1}} \otimes \rho_{B_{1}B_{2}} \otimes ...\otimes \rho_{B_{n-2}B_{n-1}} \otimes \rho_{B_{n-1}C}$ for the $n-local$ chain network captured in the fig[\ref{fig:bilocal3}]. We find for this scenario,
$$I = \frac{1}{4}\sum_{i = x,y,z}(a_{i}+a_{i}^\prime)t_{iz}^{AB_{1}}\times t_{zz}^{B_{1}B_{2}}t_{zz}^{B_{2}B_{3}}....t_{zz}^{B_{n-1}B_{n}}\times \sum_{j=x,y,z}t_{zj}^{B_{n}C}(c_{j}+c_{j}^\prime) $$ and 
$$J = \frac{1}{4}\sum_{i = x,y,z}(a_{i}-a_{i}^\prime)t_{ix}^{AB_{1}}\times t_{xx}^{B_{1}B_{2}}t_{xx}^{B_{2}B_{3}}.....t_{xx}^{B_{n-1}B_{n}}\times \sum_{j=x,y,z}t_{xj}^{B_{n}C}(c_{j}-c_{j}^\prime) $$

Calculating the value of $S_{n-local}$ and maximizing, we finally obtain, 
\begin{equation}
S_{n-local} = \sqrt{\sqrt{\Lambda_{1}\gamma_{1}\beta_{1}...}+\sqrt{\Lambda_{2}\gamma_{2}\beta_{2}...}}
\end{equation}
All the eigenvalues come from the matrix $R$ as previously discussed.

\section{Violation Criteria for star-shaped network}
\begin{figure}[h]
\centering
\includegraphics[width=12cm, height=8cm]{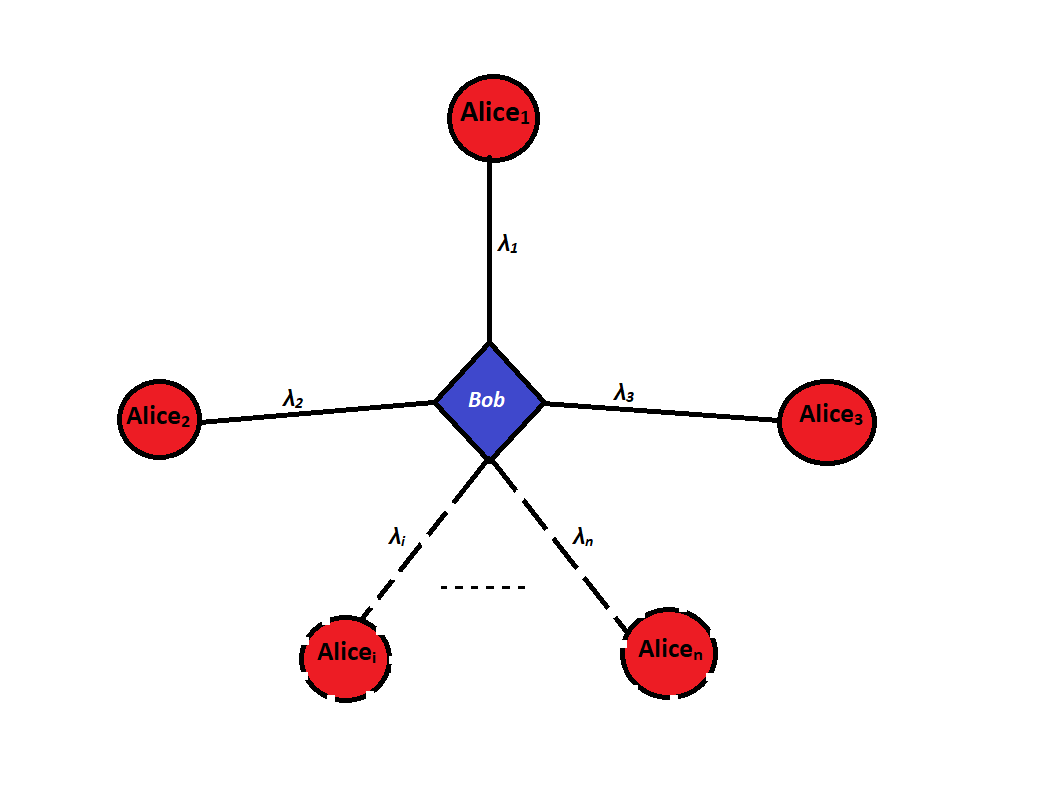}
\caption{n-local star shaped Network scenario}
\label{fig:bilocal5}
\end{figure}

In figure {\ref{fig:bilocal5}}, we consider the star-shaped network composed of n+1 parties, where a central node( referred to as Bob) shares a quantum state with each of the $n$ nodes(referred to as several Alice), i.e., states are provided by $n$ independent sources. Here, we restrict our attention to the special case where each of $n$ Alice's perform dichotomic measurements with two outputs. The inputs are denoted by $x_{i} \in \{ 0,1\}$ for each $i^{th}$ Alice and the outcomes are denoted by $a_{i} \in \{ 0,1\}$. The central node measures one input $y$ with $2^n$ outputs labeled by string $b = b^1b^2b^3...b^n$ where $b^j \in (0,1)$ for $j = 1,2,3...n $.
\\
Similar to the bilocal case, we can say, a star-shaped network scenario is $n-local$ if the joint probability distribution on the star-network can be written as,
\begin{equation}
P(a_{1}...a_{n}|x_{1}...x_{n}y) = \int (\prod_{i=1}^n d\lambda_{i}\rho(\lambda_{i})P(a_{i}|x_{i},\lambda_{i}))P(b|y,\lambda_{1}...\lambda_{n})
\end{equation}
\\
Following this, we have an inequality\cite{4} that satisfied by the above kind of probability distributions and said as a $n-local$ correlations. These inequalities are analogous to the Bell inequality in the star network scenario and they are like, 
\begin{equation}
S_{star} = \sum_{j = 1}^{2^{n-1}}I_{j}^{1/n} \leq 2^{n-2}
\end{equation}
where,
\begin{equation}
I_j = \frac{1}{2^n}\sum_{x_1...x_n}(-1)^{g_j(x_1,...,x_n)}\langle A_{x_1}^1...A_{x_n}^nB^j\rangle,
\end{equation}
each depending on a function $g_{j}(x_1,...,x_n).$ These $2^{n-1}$ functions contain an even number of $x_i$. For example, when $n$ = 2, $$g_1(x_1,x_2) = 0,$$ $$g_2(x_1,x_2) = x_1 + x_2$$
and for $n$ = 3, 
$$g_1(x_1,x_2,x_3) = 0, g_2(x_1,x_2,x_3) = x_1 + x_2,$$ $$g_3(x_1,x_2,x_3) = x_1 + x_3, g_4(x_1,x_2,x_3) = x_2 + x_3.$$
Again, for $n$ = 4,
$$g_1(x_1,x_2,x_3) = 0,  g_2(x_1,x_2,x_3) = x_1 + x_2$$  $$g_2(x_1,x_2,x_3) = x_1 + x_3, g_2(x_1,x_2,x_3) = x_1 + x_4, g_2(x_1,x_2,x_3) = x_2 + x_3$$ $$g_2(x_1,x_2,x_3) = x_2 + x_4, g_2(x_1,x_2,x_3) = x_3 + x_4, g_2(x_1,x_2,x_3) = x_1 + x_2 + x_3 + x_4$$
Now, we set the measurement choices of each Alice in a rotated bases, $(A_0^i + A_1^i) = 2\cos\alpha\vec{n}$ and $(A_0^i + A_1^i) = 2\sin\alpha \vec{n^{\prime}}$, and for the central node, Bob, we use generalized Bell State Measurements\cite{3}. The generalized n-party Bell basis ${|\psi_{r_1...r_n}\rangle}_{r_1...r_n}$, for $r_1...r_n \in {0,1}$ came from,
\begin{equation}
|\psi_{r_1...r_n}\rangle = Z^{r_1} \otimes X^{r_2}\otimes ...\otimes X^{r_n} |GHZ_n\rangle
\end{equation}
 where $$|GHZ_n\rangle = \frac{1}{\sqrt{2}}(|0\rangle^{\otimes n} + |1\rangle^{\otimes n})$$ 
We can directly jump to the generalized Bell state measurements  $B^j$, i.e., $B^j = M_0^j - M_1^j$, where,
$$M_b^j = \sum_{r_1...r_n} \delta_{b,b^j(r_1...r_n)}|\psi_{r_1...r_n}\rangle \langle \psi_{r_1...r_n}|$$

So, collecting all the elements, we can simply calculate the value of $I_j$ over a quantum state $\rho$ as,
\begin{equation}
I_j = \frac{1}{2^n}\sum (-1)^{g_j(x_1...x_n)}tr([A_{x_1}^1 \otimes ...\otimes A_{x_n}^n] \otimes (M_0^j - M_0^j)\rho)
\end{equation}
As $B^j = M_0^j - M_1^j$. Here, for $n = 3$ ,$$b^1(r,s,t) = r,  b^2(r,s,t) = r\oplus s\oplus 1,$$
 $$b^3(r,s,t) = r\oplus t\oplus 1,  b^4(r,s,t) = r\oplus s\oplus t\oplus 1,$$
 And for $n =4$ [3],
 $$b^1(r_1,r_2,r_3,r_4) = r_1, b^1(r_1,r_2,r_3,r_4) = r_2 \oplus r_3 \oplus 1, b^1(r_1,r_2,r_3,r_4) = r_2 \oplus r_4 \oplus 1,$$
 $$b^1(r_1,r_2,r_3,r_4) = r_3 \oplus r_4 \oplus 1, b^1(r_1,r_2,r_3,r_4) = r_2 \oplus 1, b^1(r_1,r_2,r_3,r_4) = r_3 \oplus 1,, b^1(r_1,r_2,r_3,r_4) = r_4 \oplus 1$$
 $$b^1(r_1,r_2,r_3,r_4) = r_2 \oplus r_3 \oplus r_4.$$
Thus, for the generic quantum state $\rho = \rho_{A_1B}\otimes...\otimes \rho_{A_nB}$, we calculate the value of $S$ from eqn(16) and maximizing the the quantity $S$ with respect to $\alpha$ and using Lagrange Multipliers, we have $$S = 2^{n-2}\sqrt{(\prod_1^n t_1^{A_i})^{\frac{1}{n}} + (\prod_1^n t_2^{A_i})^{\frac{1}{n}}}$$
 where $t_1^{A_i}$ and $t_2^{A_i}$ are the two greater eigenvalues of the correlation matrix $T_{\rho_{A_iB}}^T T_{\rho_{A_iB}}$. Here we used the argument that $\vec{n} \perp \vec{n^{\prime}}$ and $T_{\rho_{A_iB}}^T T_{\rho_{A_iB}}$ is a symmetric matrix and diagonalizable.
 
 \section{Conclusion}
 Generalizations of Bell's theorem to complex networks offer a new theoretical and experimental ground for further understanding of quantum correlations and its practical applications in information processing tasks. Similar to usual Bell-type scenarios, understanding the set of quantum correlations we can achieve and in particular, what are the optimal quantum violation of Bell inequalities, is of primal importance here. In this work, we simply generalize the result of the maximal violation of bilocal inequality to $n-local$ inequality. We set up a scenario with $n$ independent sources distributed in two different ways. First, the chain-shaped network where all parties have two hands except the two extreme parties and we obtain an analogous inequality given by Gisin\cite{1} and Andreoli\cite{14} et. al. Secondly, the $n$ independent sources are distributed in a manner that a central node connected with many other parties called star-shaped network with one input and $2^n$ outputs of the central party. Each result always provides the bilocal result when we take $n = 2$, i.e., two independent sources. So, we got a more general violation by the two-qubit states for the more general network scenario with $n-local$ inequality. It should be a possible future interest to find this bound with more efficient measurement choices and also with more complex networks.
 \section{Acknowledgement}
 The author AK acknowledges CSIR EMR-I and MKM acknowledges UGC for this work. The authors D. Sarkar and I. Chattopadhyay acknowledge it as QuST initiatives.

\end{document}